\input{psfig.sty}

\documentclass{aa} 

\usepackage{times}
\usepackage{natbib}
\usepackage{tabularx}
\usepackage{graphicx}
\usepackage{wasysym}

\begin{document}

\title{MHD evolution of a fragment of a CME core in the outer solar corona}
\author{P.Pagano\inst{{1},{2}} \and F.Reale\inst{{1},{2}} \and S.Orlando\inst{2} \and G.Peres\inst{{1},{2}}}
\institute{Dipartimento di Scienze Fisiche ed Astronomiche, Sezione di Astronomia, Universit\`a di Palermo, Piazza del Parlamento 1, 90134 Palermo, Italy
\and
INAF - Osservatorio Astronomico di Palermo, Piazza del Parlamento 1, 90134 Palermo, Italy}
   
\newcommand{\btx}{\textsc{Bib}\TeX}
\newcommand{\filename}{esapub}

\abstract
{Detailed hydrodynamic modeling explained several features of a fragment of the core of a Coronal Mass Ejection 
observed with SoHO/UVCS at $1.7$ $R_{\odot}$ on 12 December 1997, but some questions remained unsolved.}
{We investigate the role of the magnetic fields in
the thermal insulation and the expansion of an ejected fragment (cloud) traveling upwards in the outer corona.}
{We perform MHD simulations including the effects of thermal conduction and radiative losses
of a dense spherical or cylindrical cloud launched upwards in the outer corona,
with various assumptions on the strength and topology of the ambient magnetic field;
we also consider the case of a cylindrical cloud with an internal magnetic field component along its axis.}
{We find that a weak ambient magnetic field ($\beta \sim 20$) with open topology
provides both significant thermal insulation and large expansion.
The cylindrical cloud expands more than the spherical one.}
{}

\keywords{CME; MHD; Solar Corona}

\maketitle

\section{Introduction}
The study of Coronal Mass Ejections (CME) has received a strong
impulse since the launch of the Solar Heliospheric Observatory (SoHO) in 1995
\citep{Domingo1988}.  The Ultraviolet Coronal Spectrometer
(UVCS) on board SoHO \citep{Kohl1995} monitored the evolution of many CMEs and brought
detailed spectroscopic information on plasma in CMEs during their propagation
in the outer corona.  

\citet{Ciaravella2000} and \citet{Ciaravella2001} analyzed in detail a CME occurred on 12 December 1997. 
\citet{Ciaravella2001} addressed more specifically
the features observed at a distance of about $2$ $R_{\odot}$ by SoHO/UVCS,
and hydrodynamic (HD) simulations of a cloud expelled out of the corona
were performed to investigate which features of a fragment of the CME
core observed with SoHO/UVCS could be explained in terms of purely
HD effects. That model assumed the CME fragment to be already formed
and launched with an initial impulse, and studied its evolution while
traveling from the height of the EIT observation ($\sim 1.2$ $R_\odot$)
to that of the UVCS observation ($\sim 1.7$ $R_\odot$). The HD model
was able to reproduce various features observed in SoHO/UVCS UV lines
(C III and O VI). The fragment was constrained to be initially denser
than its surrounding with a density contrast $\delta n/n_0\sim 3$ and
the background density to be $\sim 10^8$ $cm^{-3}$.

From the results of HD modeling, \citet{Ciaravella2001} argued that the 
thermal insulation of the core fragment
is required to explain its evolution and to fit the UVCS
observation. However, two main questions remained open.  First, the thermal
insulation needs to be physically motivated, since the thermal conduction
is very efficient in the corona; in particular numerical simulations showed that the
CME fragment thermalizes with the hot atmosphere on a very short time scale unless the
thermal conduction is somehow suppressed. Second, the CME is
observed to expand by a factor 3 to 4 in diameter \citep{Ciaravella2000} traveling 
from the low corona to the height of the UVCS
observation, while the hydrodynamic simulations lead to a smaller expansion 
of
the fragment. Although the modeling suggests that very elongated
structures may expand more, the amount of expansion needs further investigation.

Here, we extend the modeling study described in Ciaravella et al. (2001)
and investigate whether magnetic fields may have a crucial role to answer
the questions left open in the previous study, i.e. the thermal
insulation and/or the amount of the expansion of the moving CME fragment. The
magnetic field is able to funnel the thermal conduction, and it may lead
to a strong thermal insulation of the fragment depending on the magnetic
field topology.
For this purpose, we revisit the modeling approach described in
\citet{Ciaravella2001} now including an ambient magnetic field
and using a full MHD modeling that includes non-isotropic thermal conduction,
to understand in which conditions the expelled fragment can be
thermally insulated or may expand more.
We study the evolution of the moving fragment for different topologies
and intensities of the ambient coronal magnetic field, and eventually 
consider also a fragment with an internal magnetic field, 
therefore accounting for the ejection of magnetic flux.
In section~\ref{model} we describe the model, section~\ref{results} presents
the simulations and in section~\ref{discussion}
we discuss the results.

\section{The model}\label{model}

Our scope is to study the evolution of an ejected CME core fragment as
it travels through the high corona, extending the modeling described in
Section 5.2 of Ciaravella et al. (2001). Again, we consider a simplified
model of a plasma cloud, denser than the surrounding atmosphere and moving
upwards in a stratified atmosphere with
parameters that lead to the conditions observed at $1.7$ $R_{\odot}$
with UVCS. At variance with \citet{Ciaravella2001} now we consider
the presence of an ambient magnetic field.  We model the fragment while it
travels in a $\beta \ga 1$ environment \citep{Gary2001}, where we cannot
neglect the magnetic pressure, nor the thermal one, and we include the
effect of the solar gravity, the radiative losses from an optically thin
plasma, a coronal heating (needed to keep the unperturbed atmosphere in
hydrostatic conditions), and the anisotropic thermal conduction in the
presence of a magnetic field.

For continuity with the work of Ciaravella et al. (2001), we will
first report on the evolution of a plasma cloud moving in various
configurations of ambient magnetic field and eventually consider the
case of a cloud of magnetic flux and plasma, i.e. a ``magnetized''
cloud, with an internal magnetic field.

We solve the ideal MHD equations (here in CGS conservative form):

\begin{equation}
\label{mass}
\frac{\partial\rho}{\partial t}+\vec{\nabla}\cdot(\rho\vec{v})=0
\end{equation}
\begin{equation}
\frac{\partial\rho\vec{v}}{\partial t}+\vec{\nabla}\cdot(\rho\vec{v}\vec{v})=-\nabla p+\frac{(\vec{\nabla}\times\vec{B})\times\vec{B}}{4\pi}+\rho\vec{g}
\end{equation}
\begin{equation}
\frac{\partial u}{\partial t}+\vec{\nabla}\cdot[(u+p)\vec{v}]=\rho\vec{g}\cdot\vec{v}-n^2P(T)+H-\vec{\nabla}\cdot\vec{F_c}
\end{equation}
\begin{equation}
\frac{\partial\vec{B}}{\partial t}=\vec{\nabla}\times(\vec{v}\times\vec{B})
\end{equation}

\begin{equation}
u=\frac{1}{2}\rho v^2+E
\end{equation}
\begin{equation}
\label{stato}
p=(\gamma-1)E
\end{equation}
where $t$ is the time, $\rho$ is the mass density, $n$ the number density, $p$ the thermal pressure, $T$ the temperature, $\vec{v}$ the plasma flow speed,
$E$ the internal energy, $u$ the total energy (internal plus kinetic), $\gamma$ the adiabatic index, $\vec{g}$ the gravity acceleration, $\vec{B}$ the magnetic field,
$P(T)$ the radiative losses per unit emission measure \citep{Raymond1977},
$H$ a space-dependent heating function equal to the radiative losses in the initial conditions and $F_c$ is the conductive
flux according to \citet{Spitzer1962}.
We neglect resistivity effects.

We solve numerically the equations with the MHD version of the advanced parallel FLASH code,
basically developed by the ASC / Alliance Center for Astrophysical Thermonuclear Flashes in Chicago \citep{Fryxell2000}, 
with Adaptive Mesh Refinement PARAMESH \citep{MacNeice2000}.
The MHD equations are solved using the numerical scheme HLLE ({\it Harten Lax Van Leer Einfeldt})
proposed by \citet{Einfeldt1988}.
We implemented a FLASH module to model the anisotropic thermal conduction according to \citet{Spitzer1962}.
The thermal flux is locally split into two components, along and across the magnetic field,
and is given by:

\begin{equation}
\label{thflux}
\vec{F_c}=-\kappa_{\parallel}(\nabla T)_{\parallel}-\kappa_{\perp}(\nabla T)_{\perp}
\end{equation}
where $(\nabla T)_{\parallel}$ and $(\nabla T)_{\perp}$ are the thermal gradients along and across the magnetic field,
and $\kappa_{\parallel}$ and $\kappa_{\perp}$ the thermal conduction coefficients
along and across the magnetic field lines, respectively, given by \citet{Spitzer1962}:

\begin{equation}
\kappa_{\parallel}=\delta_T\epsilon 20 \left( \frac{2}{\pi} \right)^{3/2}\frac{(k_bT)^{5/2}k_b}{m_e^{1/2}e^4Z\ln(\Lambda)}
\label{kpara}
\end{equation}
\begin{equation}
\kappa_{\perp}=\frac{8(\pi m_ik_b)^{1/2}n^2Z^2e^2c^2\ln(\Lambda)}{3B^2T^{1/2}}
\label{kperp}
\end{equation}
where $c$ is the light speed, $k_b$ is the Boltzmann constant, $m_e$ and $m_i$ are respectively the electron mass and the average ion mass,
$e$ is the electron charge, $Z$ is the average atomic number,
$\delta_T$ and $\epsilon$ parameters depends on the chemical composition, and in a proton-electron plasma:
$\delta_T=0.225$ and $\epsilon=0.419$;
$\ln(\Lambda)$ is the Coulomb logarithm which is (for the coronal plasma) $\ln(\Lambda)\sim20$.

The unperturbed corona is assumed to be in magnetohydrostatic conditions, isothermal ($T=1.5\times10^6$ K)
and stratified by gravity with a number density $n_0=10^8$ $cm^{-3}$ at the initial height of the cloud centre.
The cloud, assumed in pressure equilibrium with the atmosphere, is denser ($n_{CME}=4n_{ATM}$) and colder than its surrounding
and its center is a height of $\sim 0.15$ $R_{\odot}$ above the photosphere.
It is also stratified to preserve initial isobaric equilibrium.
As in \citet{Ciaravella2001}, the cloud at $t=0$ s is assumed circular,
which
in a 2-D cartesian geometry is equivalent to a cylindrical
cloud with infinite $y$-extension in a proper 3-D domain. 

For all the simulations, the radius of the cloud is $5\times10^{9}$ cm
and the initial upward velocity of the cloud is 400 km/s,
supersonic but smaller than the escape velocity ($\sim 600$~km/s),
and large enough to reach the height of the UVCS observation at $1.7$ $R_{\odot}$.
The simulations describe the evolution of the plasma for a time interval of 3000~s,
i.e. the time taken by the cloud to reach the UVCS observation height.
The coordinate system is 2-D cartesian ($x$,$z$),
the vertical direction is the $z$ coordinate. 
The computational domain extends from the solar photosphere ($z=0$ cm) to
$\approx R_{\odot}$ ($z=7\times10^{10}cm$) in the $z$ direction
and to $\approx 0.4R_{\odot}$ in the $x$ direction ($x=0$ cm to $x=3\times10^{10}cm$).
The symmetry of the system along the $x=0$ cm axis allows us to model half spatial domain.
The adaptive mesh algorithm yields an effective resolution of
$\sim3\times10^8$ cm (16 zones per cloud radius).
Reflection boundary conditions are imposed along the $x=0$ cm axis consistently with the symmetry.
At the upper and external boundaries we set zero-gradient (outflow) boundary conditions.
Fixed values are imposed at the lower boundary ($z=0$ cm).
We use 2-D geometry for most of our modeling.

\begin{table*}
\caption{Numerical simulations}
\label{tab:sim}
\begin{center}
\renewcommand{\arraystretch}{0.8}
\begin{tabular}{l c c c c c c c}
\hline
\hline
Name & Magn. & B dipole & $\beta$ & Cloud & Cloud & Thermal & Geometry \\
 & field (B) & topology &  & shape & internal $B_y$ & conduction & \\
\hline
HC & No & -- & -- & Cylinder & -- & No & 2-D cartesian (x, z) \\
HS & No & -- & -- & Sphere & -- & No & 2-D cylindrical (r, z) \\
MCLC & Yes & Closed & Low ($\sim 0.25$) & Cylinder & -- & Yes & 2-D cartesian (x, z) \\
MOLC & Yes & Open & Low ($\sim 0.25$) & Cylinder & -- & Yes & 2-D cartesian (x, z) \\
MCHC & Yes & Closed & High ($\sim 25$) & Cylinder & -- & Yes & 2-D cartesian (x, z) \\
MOHC & Yes & Open & High ($\sim 25$) & Cylinder & -- & Yes & 2-D cartesian (x, z) \\
MOHS & Yes & Open & High ($\sim 25$) & Sphere & -- & Yes & 3-D cartesian (x, y, z) \\
MOHCL & Yes & Open & High ($\sim 25$) & Cylinder & Low (0.2 G) & Yes & 2-D cartesian (x, z) \\
MOHCH & Yes & Open & High ($\sim 25$) & Cylinder & High (1 G) & Yes & 2-D cartesian (x, z) \\
\\
\hline
\end{tabular}
\end{center}
\end{table*}

\begin{figure}[!htcb]
\centering
\includegraphics[scale=0.72,clip,viewport=35 290 300 558]{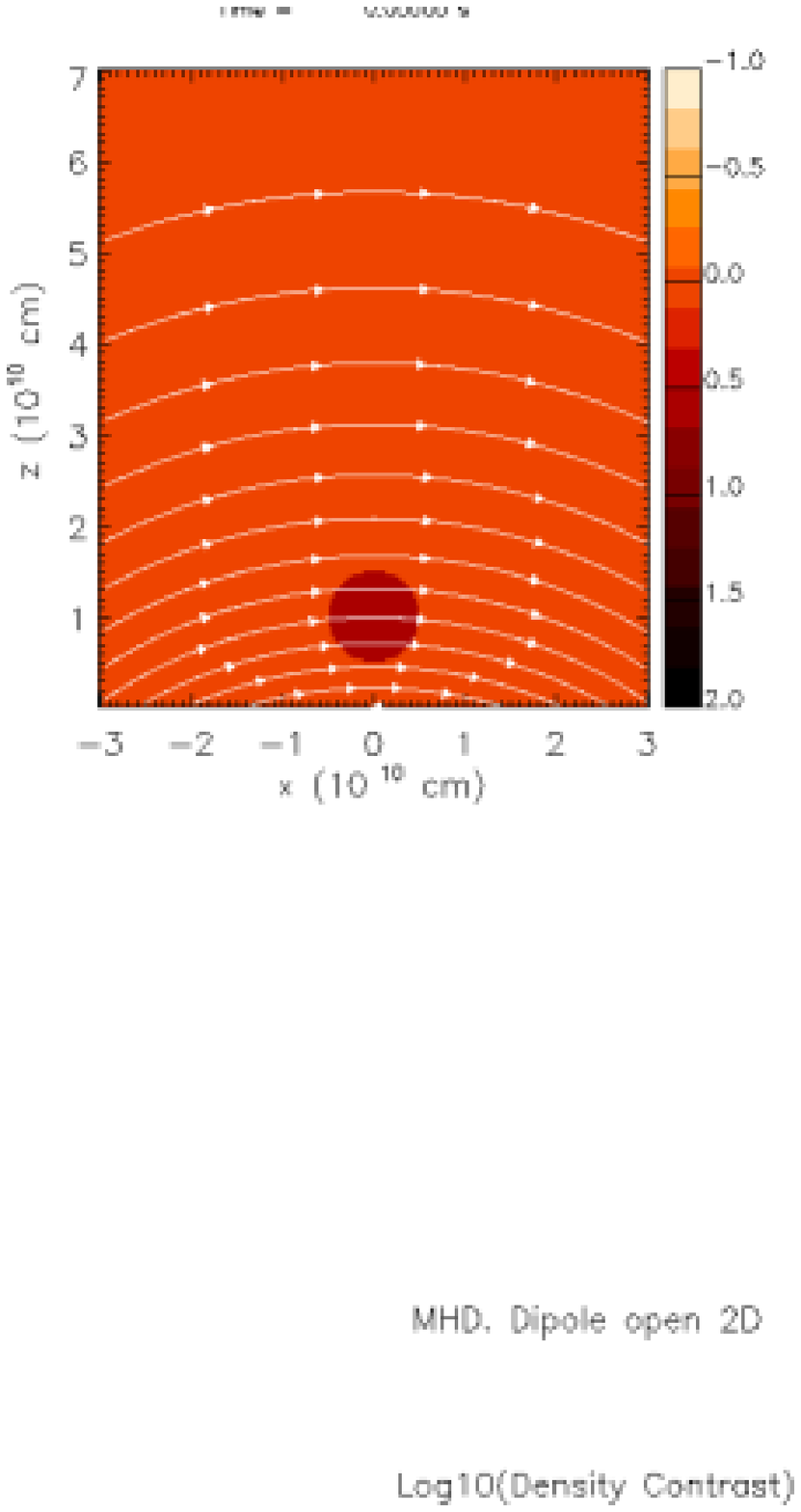}
\includegraphics[scale=0.72,clip,viewport=35 290 300 558]{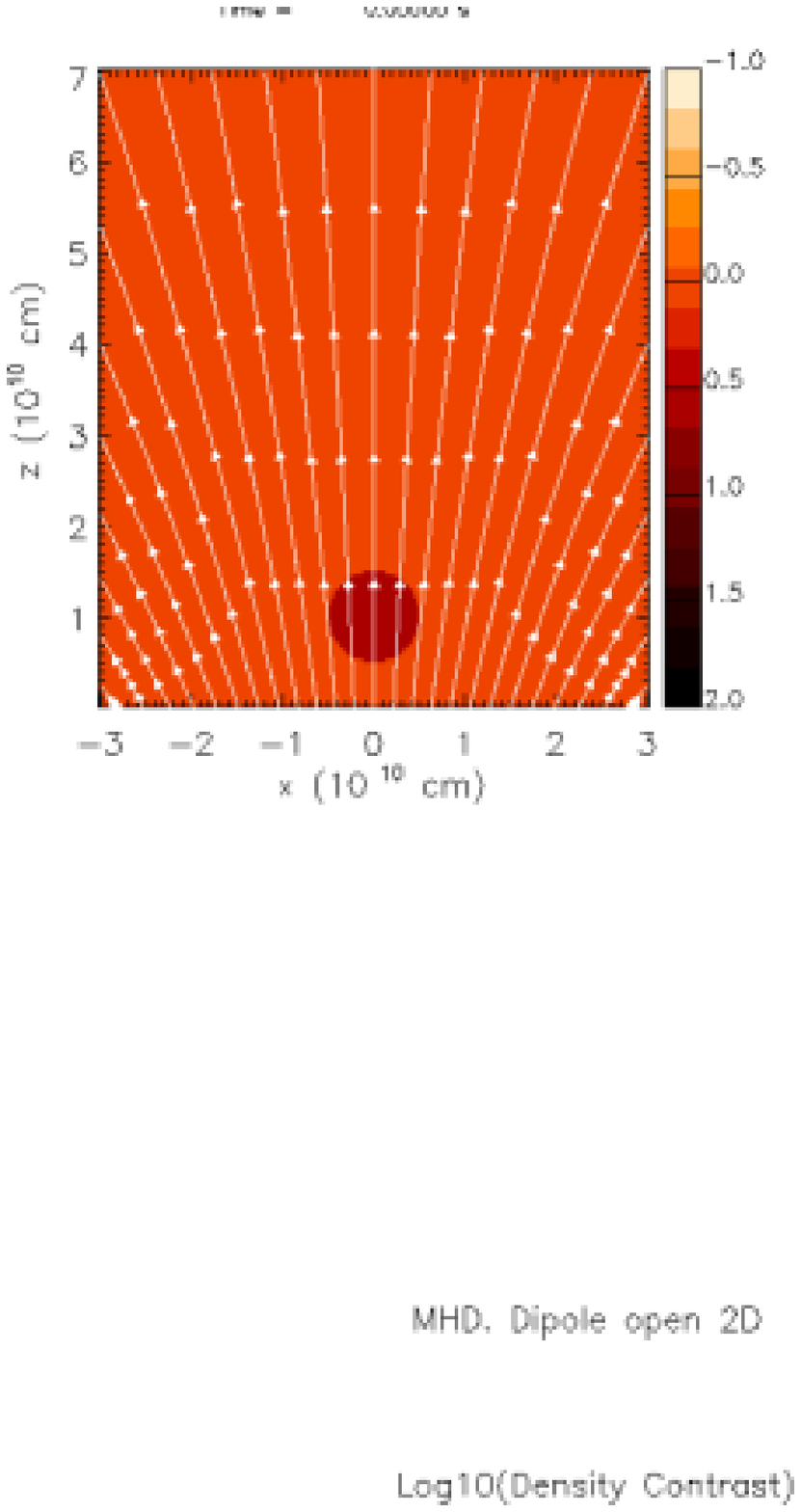}
\caption{Magnetic field lines and color maps of the density contrast ($\rho/\rho_0$) 
for two different initial configurations of magnetic 
field:
{\it closed} field ({\it upper panel}, those starting with MC in Table~\ref{tab:sim}) and {\it open} field ({\it lower panel}, those starting with MO in Table~\ref{tab:sim}).
}
\label{ci}
\end{figure}

From this setup we perform a basic set of simulations with different ambient 
coronal magnetic fields, $\vec{B}$, in which the cloud evolves, i.e. (a) with
no magnetic field,
(b) with a dipolar magnetic field on the equatorial plane ({\it closed} field), 
(c) with a dipolar magnetic field on the polar axis ({\it open} field).
Since we follow the cloud evolution much later than the launch and far from
the origin site, our ambient magnetic field is relatively weak, as of
the outer corona \citep{Gary2001}. In particular, we assume that the
dipole lays in the center of the Sun and its strength is tuned so to have
$\beta\sim 25$ at the cloud height, as basic value.
Figure \ref{ci} shows the initial conditions for the simulations 
with an ambient dipolar field ({\it closed} and {\it open}). In such
configurations, the highest $\beta \approx 40$ is at the base of the
atmosphere (where the pressure is higher), $\beta$ varies from 30 to
20 inside the initial cloud, and then it settles to $\beta \sim 15$
from a height of 1.5 $R_\odot$ up.
Simulations with more strong magnetic fields have also been considered
(see below), and there the fractional variation of $\beta$ is the same
throughout the spatial domain.

\begin{figure}[!htcb]
\centering
\includegraphics[scale=0.35,clip]{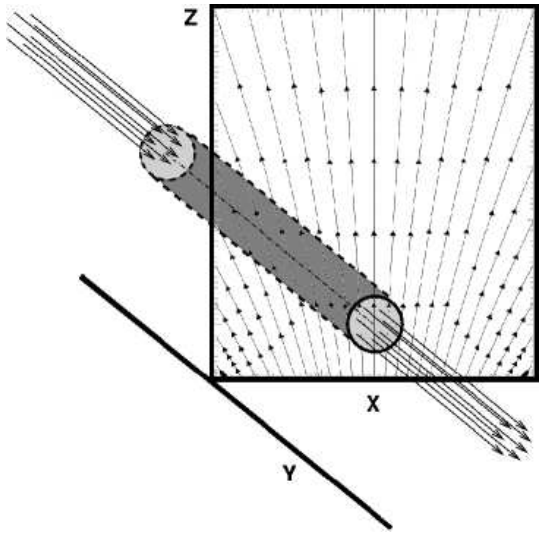}
\caption{Sketch of the initial configuration of the simulations with an open dipolar ambient magnetic field and 
an initially magnetized cloud (MOHCL and MOHCH in Table~\ref{tab:sim}). The ambient dipolar magnetic field is shown in the x-z plane 
(see lower panel of Fig.~\ref{ci}).
The additional magnetic field component is set inside the cloud (the long cylinder) along the $y$-direction.}
\label{fig:magn_cloud}
\end{figure}

Other simulations are also performed: we repeat simulations with no
magnetic field and with open dipole magnetic field for a 
spherical cloud (instead of cylindrical),
the same as presented in \citet{Ciaravella2001};
we make some simulations with a stronger ambient magnetic field (i.e. $\beta\sim 0.25$ at the height
of the centre of the initial cloud),
a magnetic-field-dominated initial regime.
To account for the ejection of both magnetic flux and plasma
we replicate the simulation with the open dipole ambient magnetic field for an initially magnetized cylindrical cloud, i.e.
an additional magnetic field component inside the cloud along $y$
(i.e. along the axis of the cloud), as shown in Fig.~\ref{fig:magn_cloud}. We 
consider two different values of $B_y$: $B_y = 0.2$~G and $B_y = 1$ G.
With $B_y=0.2$~G, the cloud parameters are left unchanged, therefore creating a small overpressure on the ambient
medium. With $B_y=1$~G, the cloud temperature is set equal to 40,000 K
(i.e. much lower than the reference value) to preserve the initial
condition of isobaric cloud.

Table~\ref{tab:sim} lists the relevant simulations presented in this work,
identified by the presence or the absence of magnetic field, 
the magnetic field topology, 
the value of $\beta$ at the initial height of the cloud centre,
the cloud shape, 
the magnetic field inside the cloud, 
the presence or absence of thermal conduction, 
the geometry of the computational domain.

The simulation with open magnetic field and a spherical cloud requires 3-D modeling and
we extend the cartesian domain also in the $y$-direction to $y=3\times10^{10}$ cm.
We have performed these simulations on a {\it Cluster Linux EXADRON} with 24 {\it Opteron 250 AMD} processors
and on a {\it IBM SP Cluster} with 512 {\it IBM Power5} processors.
The 2-D simulations have required $\sim300$ hours of computational time and the 3-D $\sim3000$ hours.

\section{Results}\label{results}
\subsection{Cylindrical cloud}
\subsubsection{Simulations with no magnetic field}
\begin{figure}[!htcb]
\centering
\includegraphics[scale=0.47,clip,viewport=60 145 377 540]{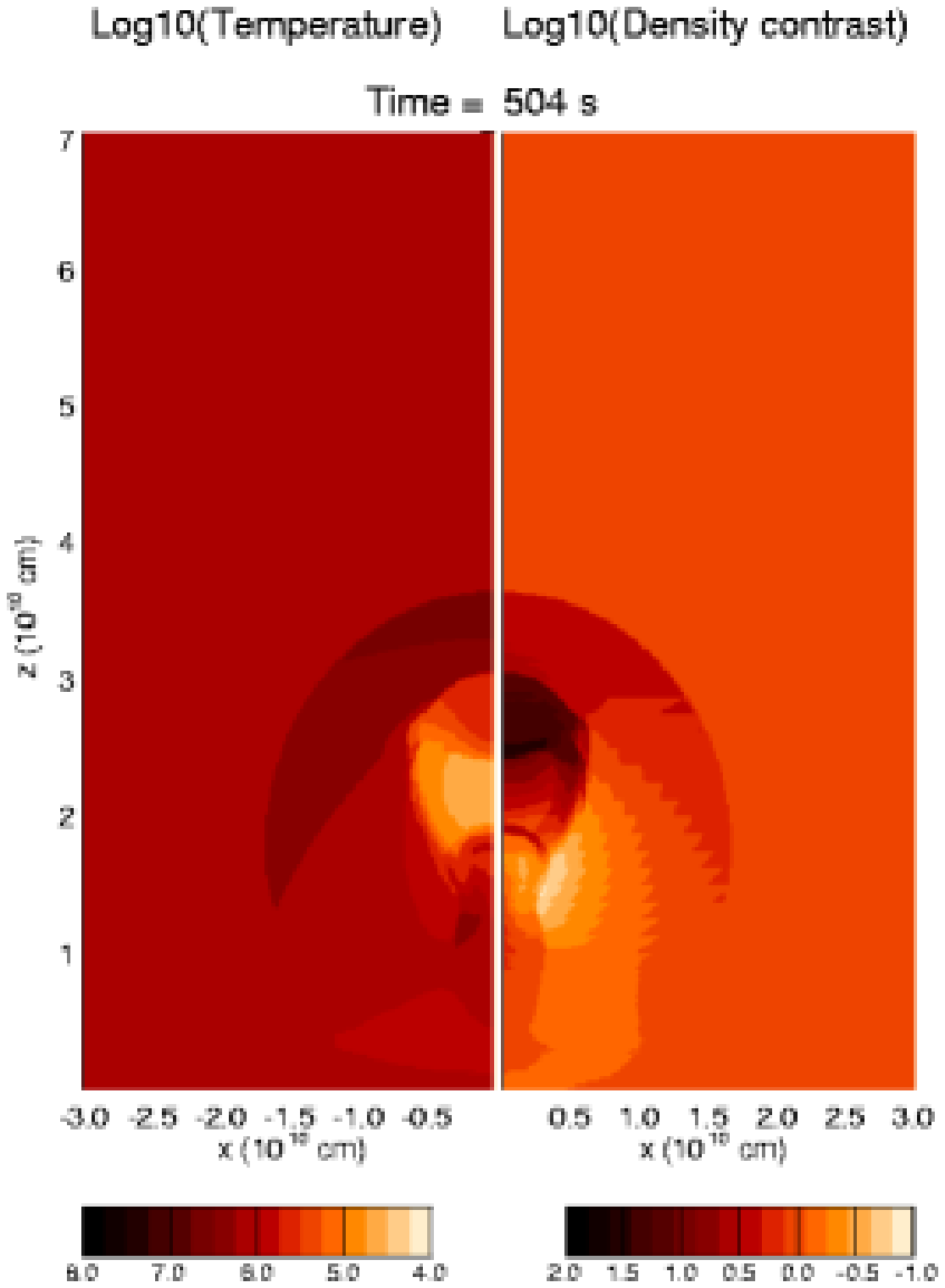}
\includegraphics[scale=0.47,clip,viewport=60 145 377 520]{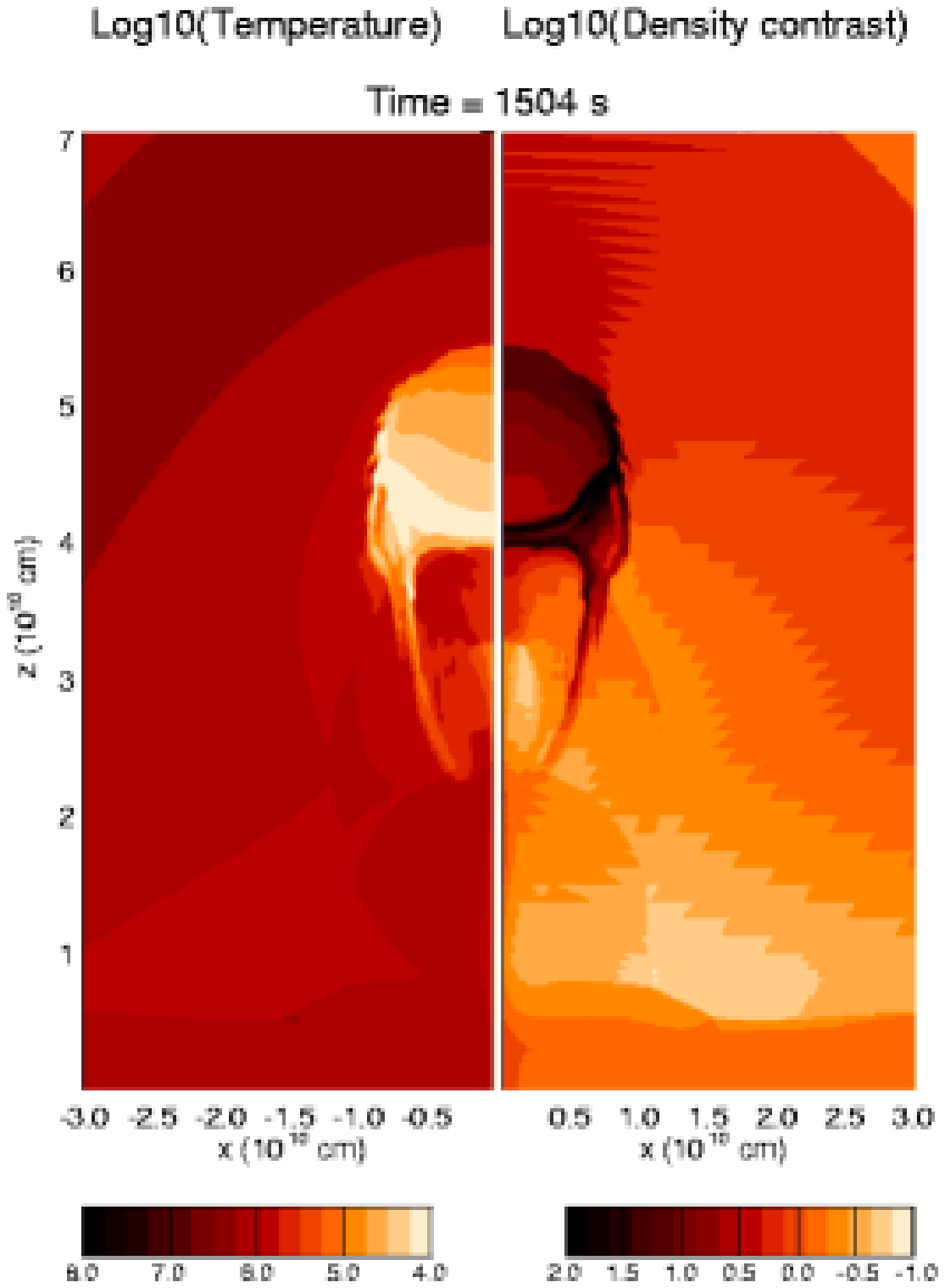}
\includegraphics[scale=0.47,clip,viewport=60 105 377 520]{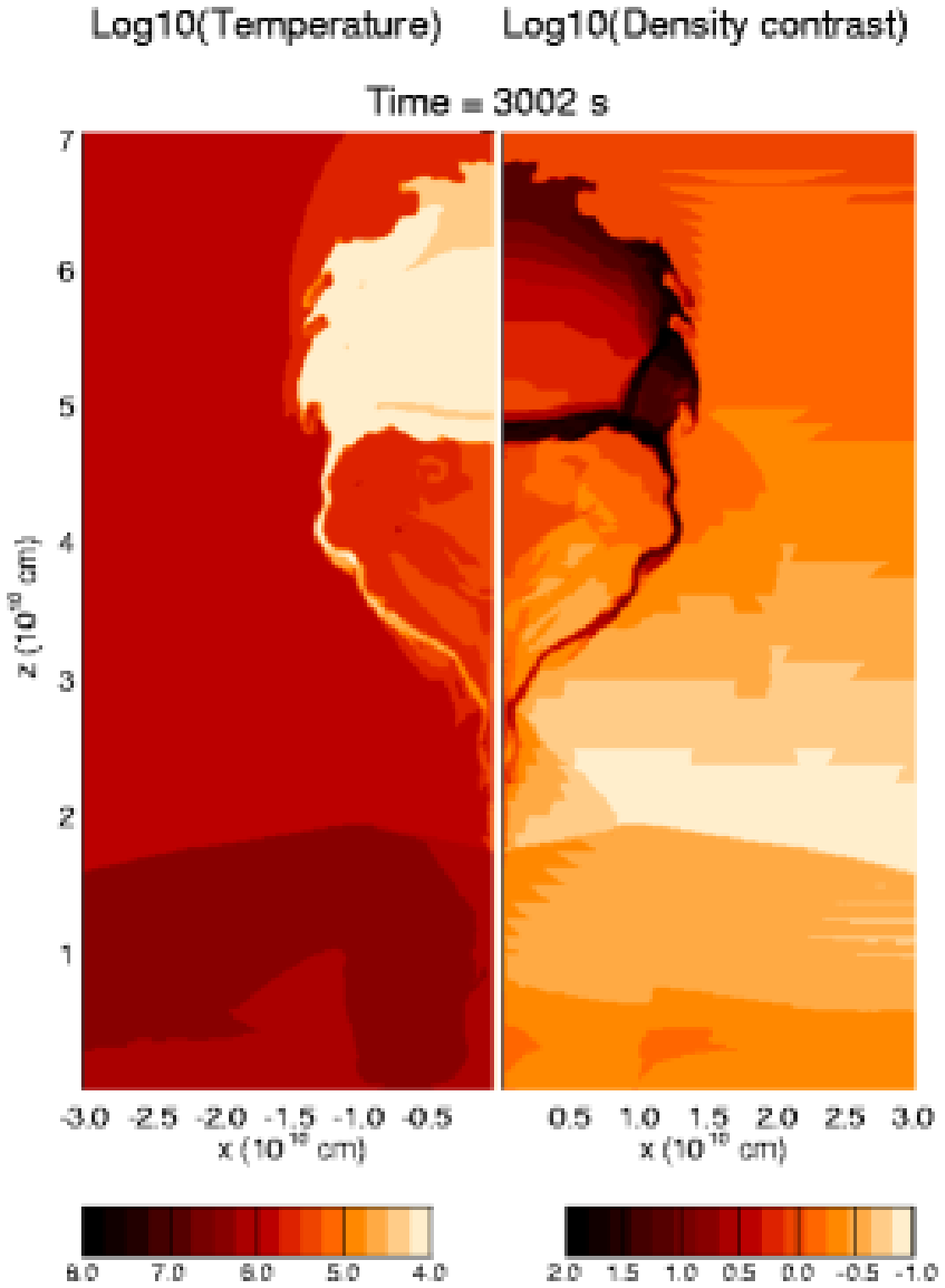}
\caption{Color maps of temperature ({\it left of each panel}) and density contrast, 
$\rho/\rho_0$ ({\it right of each panel})
at times $t = 500$ s, 1500 s and 3000 s, for the simulation with
no magnetic field (and no thermal conduction, HC in Table~\ref{tab:sim}).}
\label{HDcartdenstemp}
\end{figure}

We start the simulation with a cylindrical cloud in the environment without
magnetic field as the basic model (case HC, abbreviation for {\it Hydrodynamic model, Cylindrical 
cloud}, in Table~\ref{tab:sim}).
As in \citet{Ciaravella2001}
the thermal conduction is set to zero in this simulation.
Figure \ref{HDcartdenstemp} shows color maps of the density
contrast (ratio of the density to the density of the static atmosphere
$\rho/\rho_0$)
and the temperature at times $t=500$ s, $1500$ s,and $3000$ s, i.e. at two
intermediate times and at the final time.

The cloud initial velocity is higher than the local sound speed ($c_S\sim 200$~km/s) 
and a shock front soon departs radially from the cloud.  While moving
upwards, the cloud expands and dynamic instabilities develop at its
boundary, changing its shape and forming small scale structures,
departing from it.  At the final time the cloud has evolved into a cold
extended core with two thin tails and has mostly lost memory of its
initial circular shape.  The core becomes colder and denser during the
evolution because of the radiative losses and of the absence of thermal
conduction.  Since the cloud is stratified, thermal instabilities first
occur in the lower (and denser) arc-shaped part, and at the end
they involve the whole cloud. 
Because of the relative motion between
the cloud and the atmosphere, Kelvin-Helmholtz and Rayleigh-Taylor
instabilities develop in a time-scale \citep{Chen1972}:

\begin{equation}
\label{KHRT}
\tau_{KHRT}\approx\frac{\gamma}{\gamma-1}\frac{c_S}{g}\sim10^3~s
\end{equation}
where $c_s$ is the sound speed.
The motion of the cloud is mostly ballistic, although it loses part
of its mechanical energy (i.e. kinetic plus potential gravitational)
in the hydrodynamic interaction with the atmosphere, and at $3000$ s
the cloud has practically stopped.
To quantify the expansion of the
cloud while it moves upwards, we define an expansion factor ($F_e$) as
the ratio between the maximum width of the cloud (in the $x$-direction) 
and its initial radius, 
which can be compared to the factor 3-4 estimated by \citet{Ciaravella2000}.
At the end the expansion stops and the pressure equilibrium between the
cloud and its surroundings is recovered.  We obtain $F_e \approx 2.9$
at t=3000 s for this simulation.

We do not present here hydrodynamic simulations
with the thermal conduction. \citet{Ciaravella2001} showed that if 
the thermal conduction were not suppressed, 
the core would mostly evaporate because of heating by 
the hot surrounding corona and would shrink to a very small cold
knot.

\subsubsection{Simulations with magnetic field}\label{Magneticfield}

We now report on results of simulations with an initially cylindrical cloud, 
with non-zero ambient magnetic field,
and with the thermal conduction strongly effective along the
magnetic field lines
(MCLC, abbreviation for {\it Magnetohydrodynamic model, Closed field, Low $\beta$, Cylindrical cloud}, 
MOLC, MCHC, and MOHC in Table~\ref{tab:sim}).

\begin{figure}[!htcb]
\centering
\includegraphics[scale=0.47,clip,viewport=60 105 377 540]{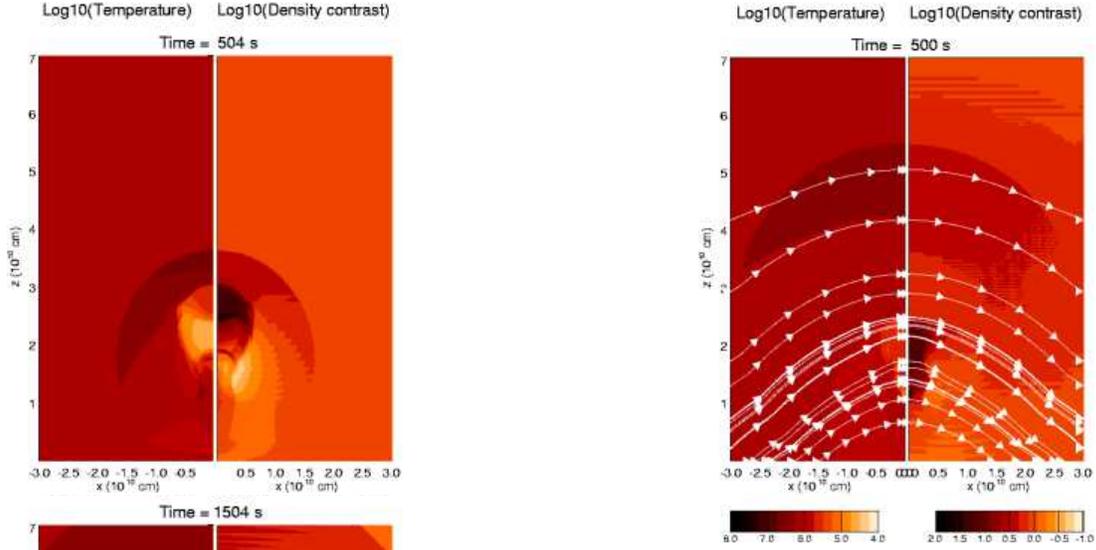}
\caption{Magnetic field lines and color maps of temperature and density contrast at time $t=500$~s 
for the simulation with a strong {\it closed} magnetic field (MCLC in Table ~\ref{tab:sim}).
The magnetic field strength is $\approx 2.7$ G at $z\sim1.5\times10^{10}$ cm along the x=0 axis (high density field lines) and
no field line was drawn in the region where it becomes less than $\sim 1$ G ($z>5\times 10^{10}$ cm).}
\label{dc2dens}
\end{figure}

In the low $\beta$ simulations (MCLC, MOLC) the cloud does not show significant evolution,
because the magnetic field is too strong to be perturbed by the
moving cloud. The strong magnetic tension does not allow the cloud to move upwards, either.
As an example, figure \ref{dc2dens} shows the case of the
{\it closed} dipolar field (MCLC) at $t=500$ s, with $\beta\sim0.25$ at the initial height of the cloud centre.

\begin{figure}[!htcb]
\centering
\includegraphics[scale=0.47,clip,viewport=60 145 377 540]{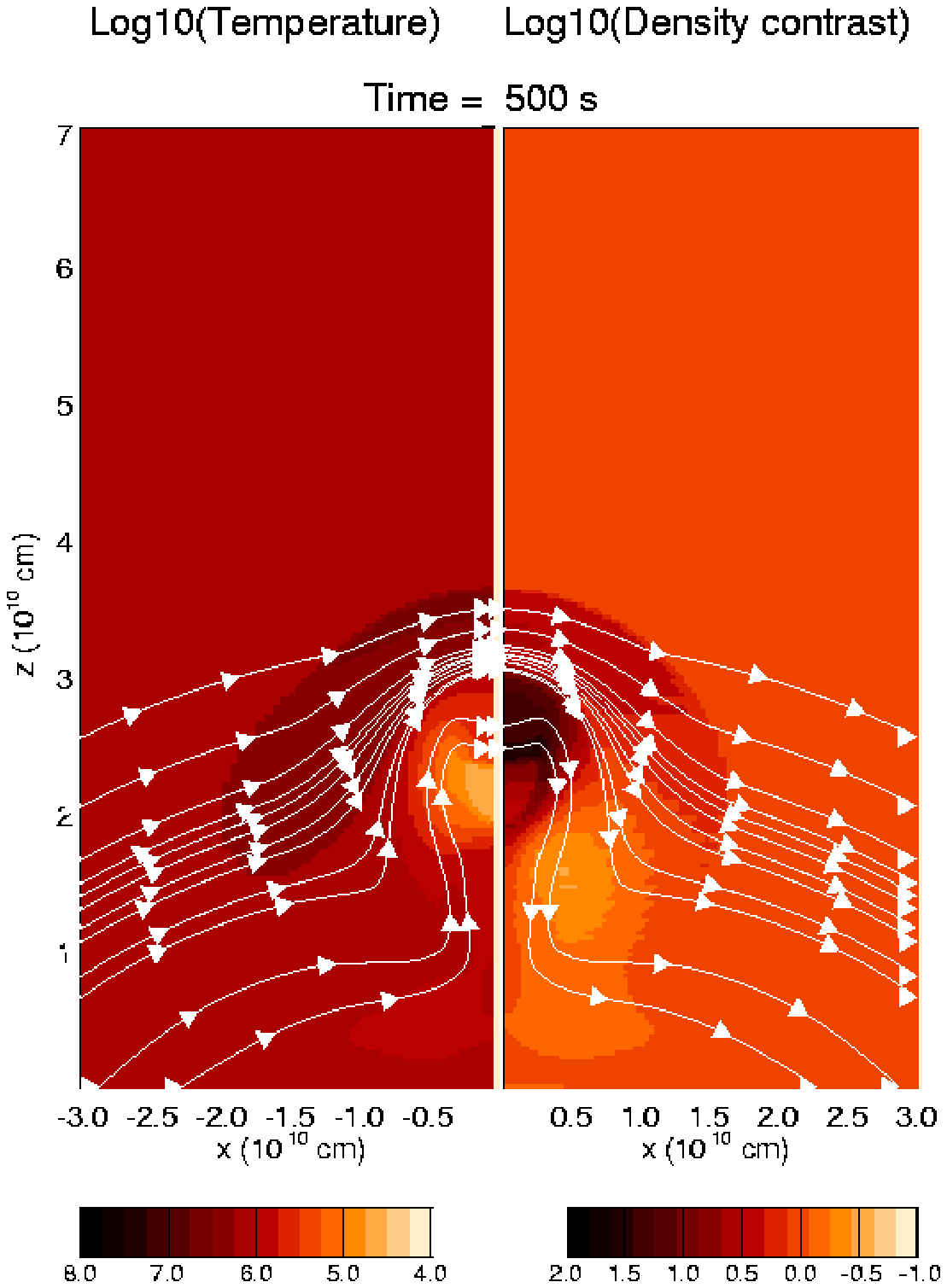}
\includegraphics[scale=0.47,clip,viewport=60 145 377 520]{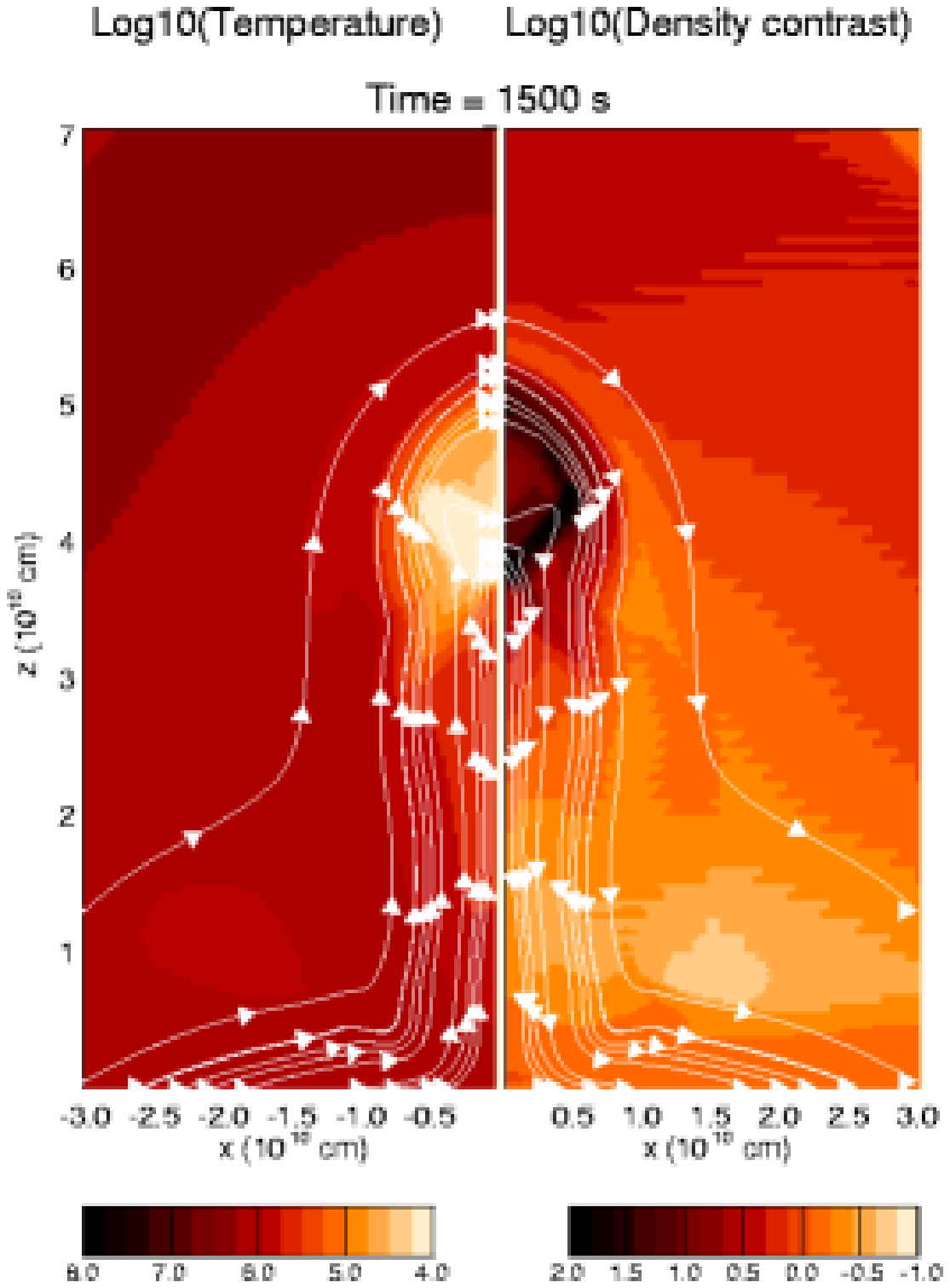}
\includegraphics[scale=0.47,clip,viewport=60 105 377 520]{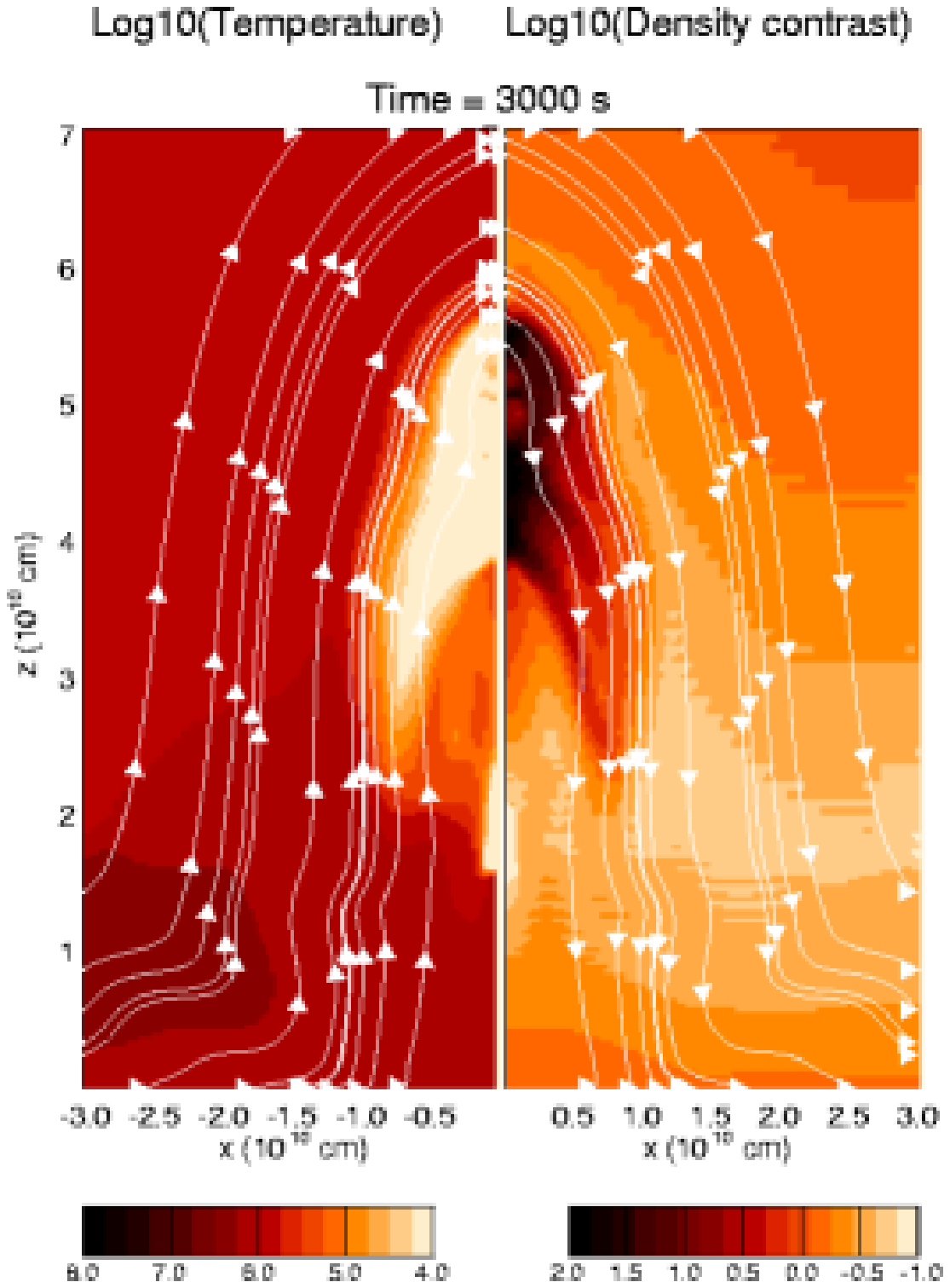}
\caption{Magnetic field lines and color maps of temperature and density contrast at t=500 s, 1500 s and 3000 s
for the simulation with weak {\it closed} magnetic field (MCHC in Table~\ref{tab:sim}).
For reference, at t=500 s the strength of the magnetic field ranges between $\approx0.05$ G 
(in the upper part of the domain where no magnetic field lines are drawn) and $\approx 0.7$ G (just above the
cloud where the lines are very dense).}
\label{dc02dens}
\end{figure}

In the high $\beta$ ambient medium, instead, the cloud moves upwards and expands in the outer corona.
Figure \ref{dc02dens} shows the evolution of the density contrast and the temperature with color maps at $t=500$~s, 1500~s and 3000~s
for the simulation with the {\it closed} dipolar field for $\beta\sim25$ at the initial cloud height.
Since the initial cloud velocity (400 km/s) is higher than both the sound speed and the Alfven speed ($v_A \sim 10^7$ cm/s),
a fast MHD shock propagates radially from the cloud.
During the evolution no hydrodynamic instability develops
because of the suppression by the thermal conduction and the magnetic field.
In fact, they are effective on time scales smaller than $\tau_{KHRT}$ (Eq.~\ref{KHRT}), 
i.e. of the order of, respectively:
\begin{equation}
\label{tcond}
\tau_{c}=\frac{21 n k_b L^2}{2 k_0 T^{2.5}}\sim 100~s
\end{equation}
\begin{equation}
\label{talf}
\tau_{a}=\frac{L}{v_A}\sim 100 ~ s
\end{equation}
\noindent
where $k_0T^{2.5}=k_{\parallel}$ and for typical coronal density with $k_0=9.2\times10^{-7}$ (c.g.s. units, from Eq.\ref{kpara}),
$v_A$ the Alfven speed, $L$ a characteristic length scale, which we have conservatively taken as
half of the cloud radius.
The cloud drags the magnetic field lines which gradually envelope the cloud and
inhibit thermal exchanges between the cloud and the surrounding medium.
The consequent strong thermal insulation leaves the plasma free
to cool down to temperatures below $T\approx10^{5.6}$ K at which it emits the UV lines of interest.
The upward motion of the cloud is braked by the restoring forces of the magnetic field;
only the upper part of the cloud reaches the UVCS observation height.
The cloud expansion is limited by the confining effect of the magnetic field, which becomes
up to 5 times stronger just outside the cloud ($\beta \sim 0.1$) than inside it.
At the end, the expansion factor of the cloud is $F_e \approx 2.1$ at $t\sim3000$ s 
(i.e. $\sim70\%$ of the reference case).

\begin{figure}[!htcb]
\centering
\includegraphics[scale=0.47,clip,viewport=60 145 377 540]{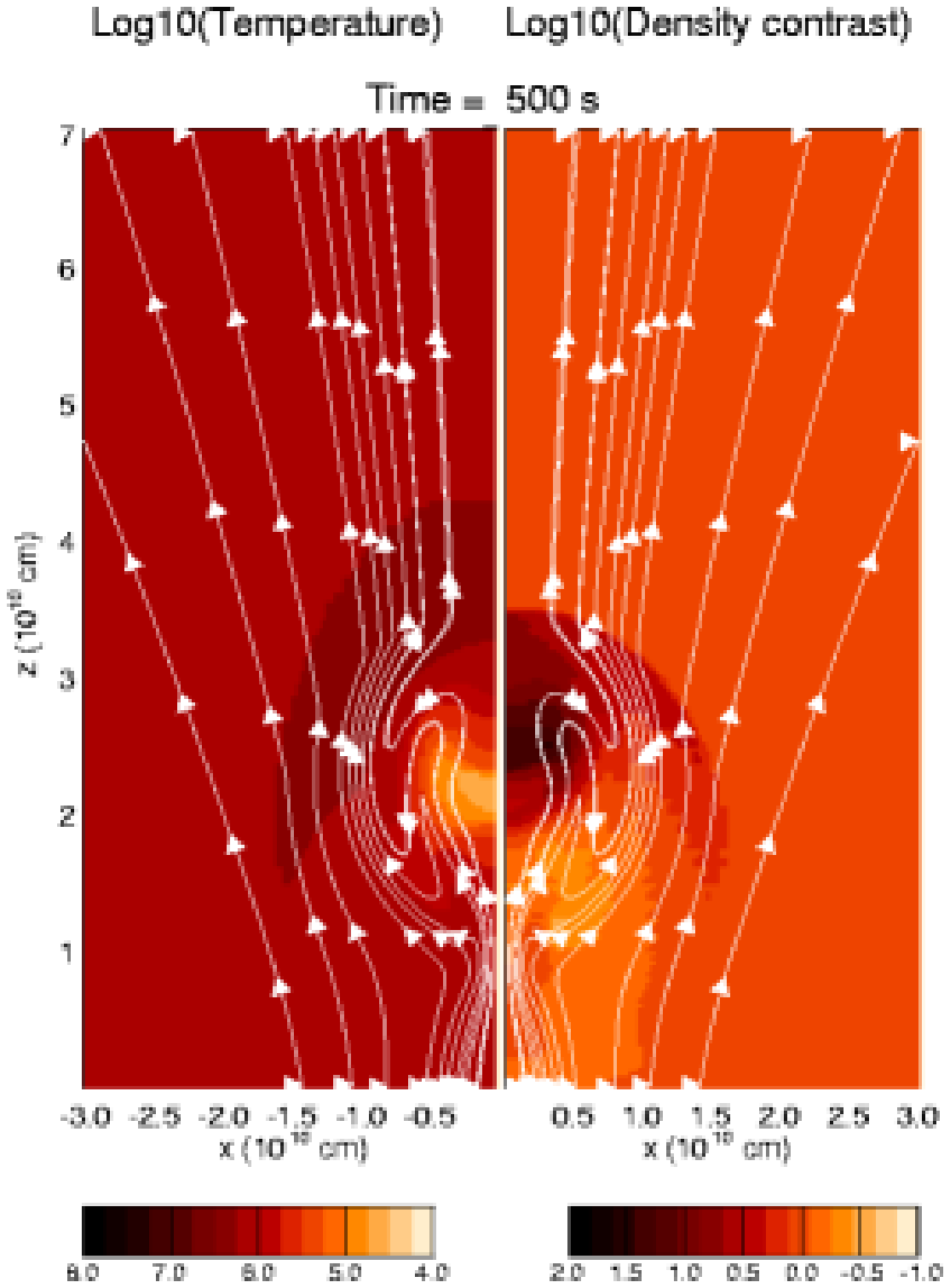}
\includegraphics[scale=0.47,clip,viewport=60 145 377 520]{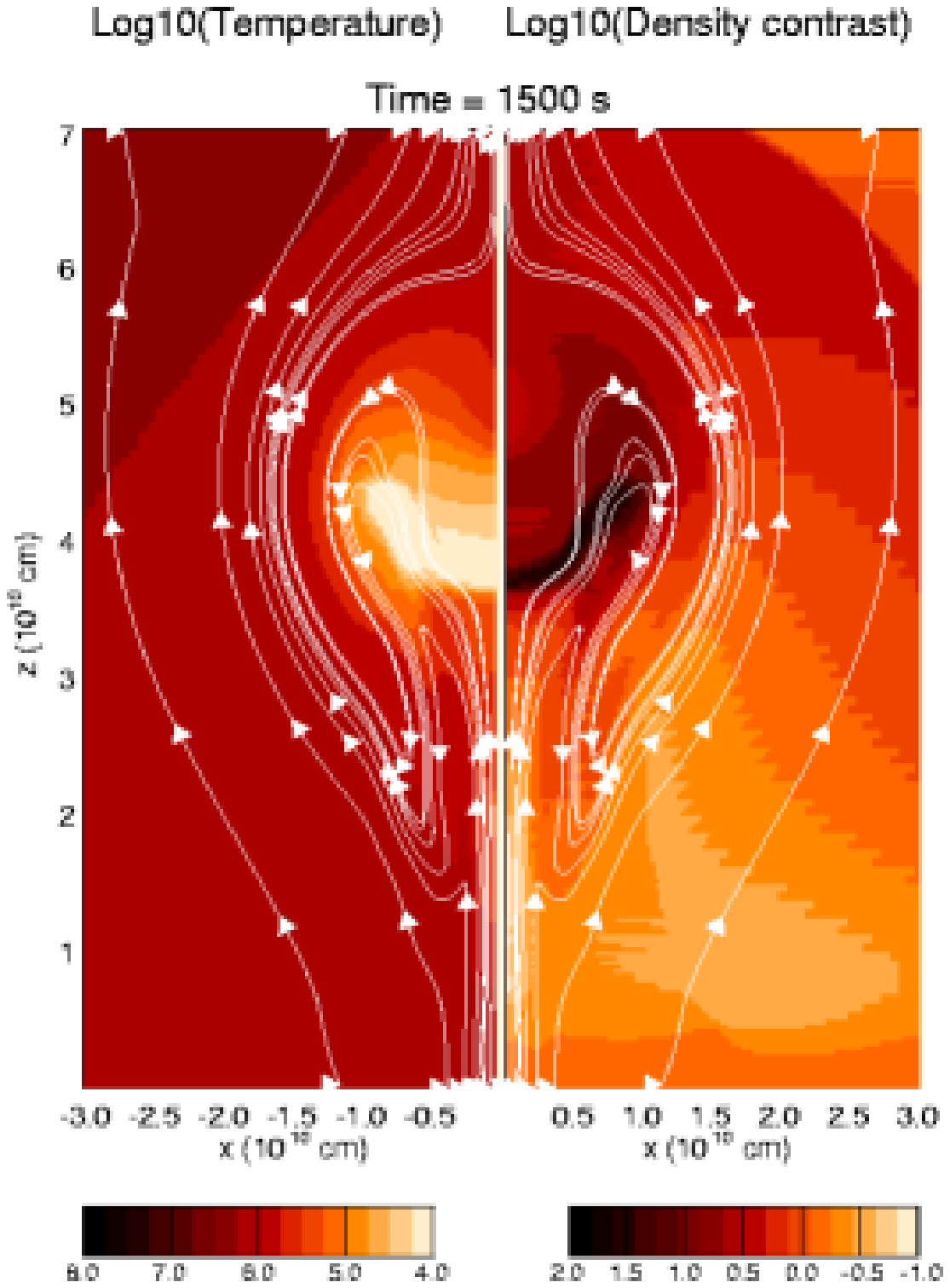}
\includegraphics[scale=0.47,clip,viewport=60 105 377 520]{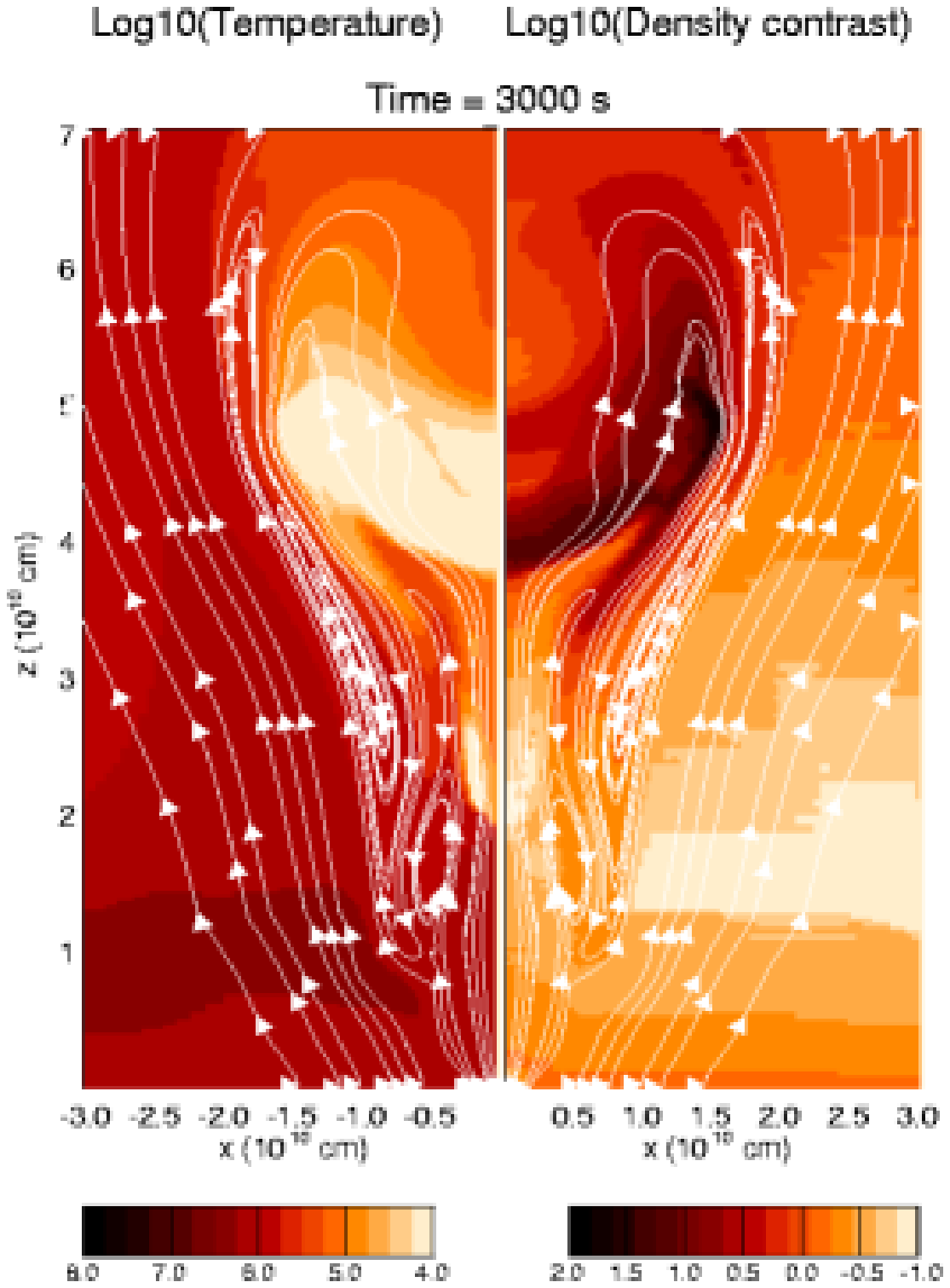}
\caption{Magnetic field lines and color maps of temperature and density contrast at t=500~s, 1500~s and 
3000~s for the simulation with weak {\it open} magnetic field (MOHC in Table~\ref{tab:sim}).
The strength of the magnetic field ranges from $\approx0.03$ G to $\approx0.5$ G.}
\label{do02dens}
\end{figure}
The cloud travels much longer distances when it moves in an {\it open} ambient magnetic field.
Figure \ref{do02dens} shows the evolution of the density contrast and of the
temperature at $t=500$ s, $1500$ s and $3000$ s for the related simulation.
Again, a fast MHD shock propagates upward radially from the cloud and no hydrodynamic instability develops.
At variance with the {\it closed} field case, in the {\it open} field the cloud moves ballistically to the outer corona.
The cloud is thermally insulated laterally, because the initial field direction is mostly maintained during the evolution.
In the upper corona, the magnetic field is weak and, therefore, perturbed by the rising cloud (at $t>2000$ s).
A strong magnetic field horizontal component develops ahead of the cloud and inhibits thermal conduction with the region above the cloud.
As the magnetic field in the cloud is frozen and the cloud moves fast with respect to the Alfv\'en scale times, it drags the magnetic field lines,
and inversely-directed downward magnetic field components and current sheets are produced near the cloud.
The {\it open} magnetic field naturally favours the cloud expansion and, in fact,
the final expansion factor is $F_e \approx 3.9$ at $t\sim3000$ s (i.e. $\sim35\%$ more than the reference case).

\subsection{Spherical cloud}\label{Spherical cloud}

\begin{figure}[!htcb]
\centering
\includegraphics[scale=0.40,clip,viewport=60 145 377 540]{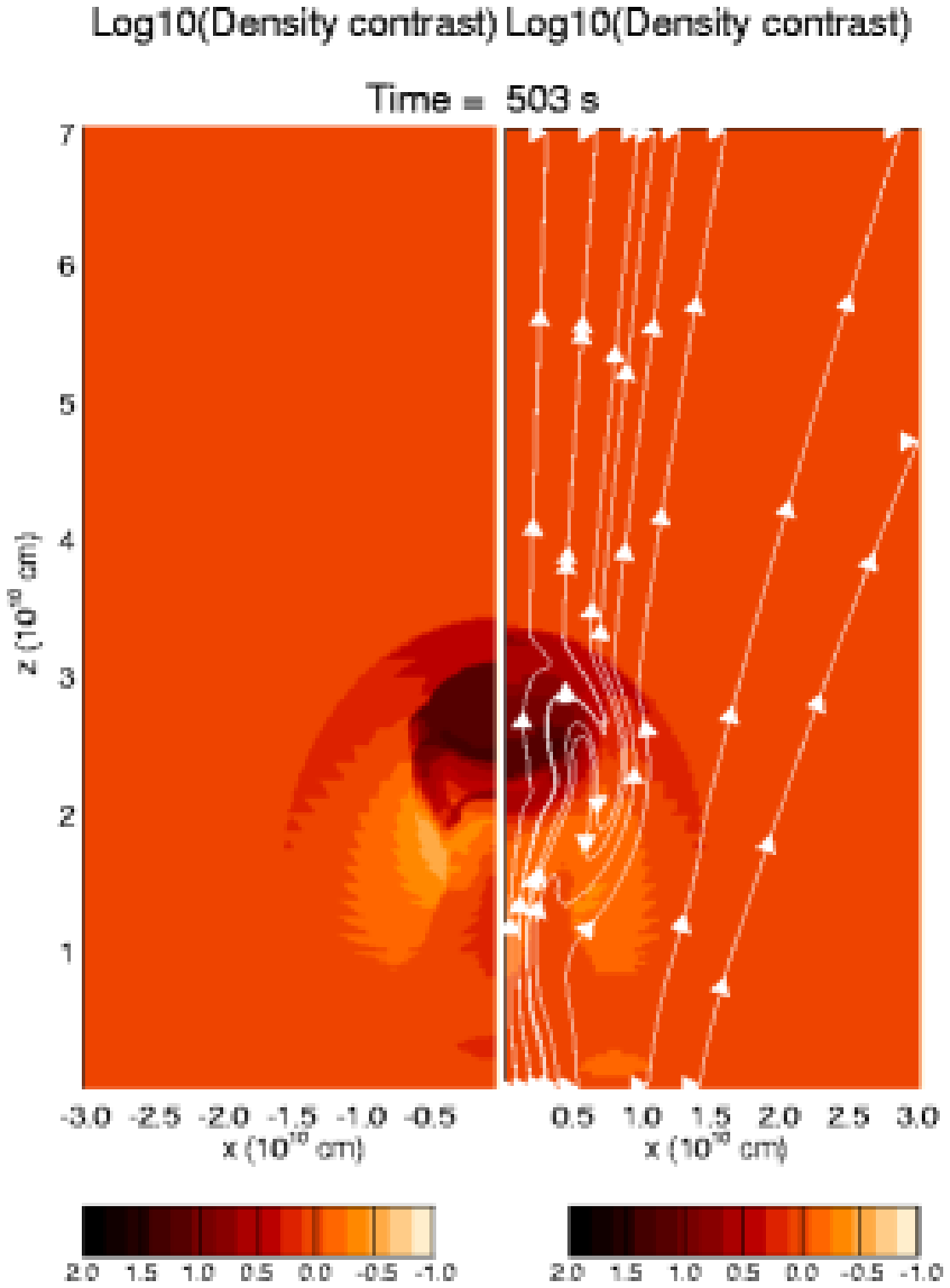}
\includegraphics[scale=0.40,clip,viewport=60 145 377 520]{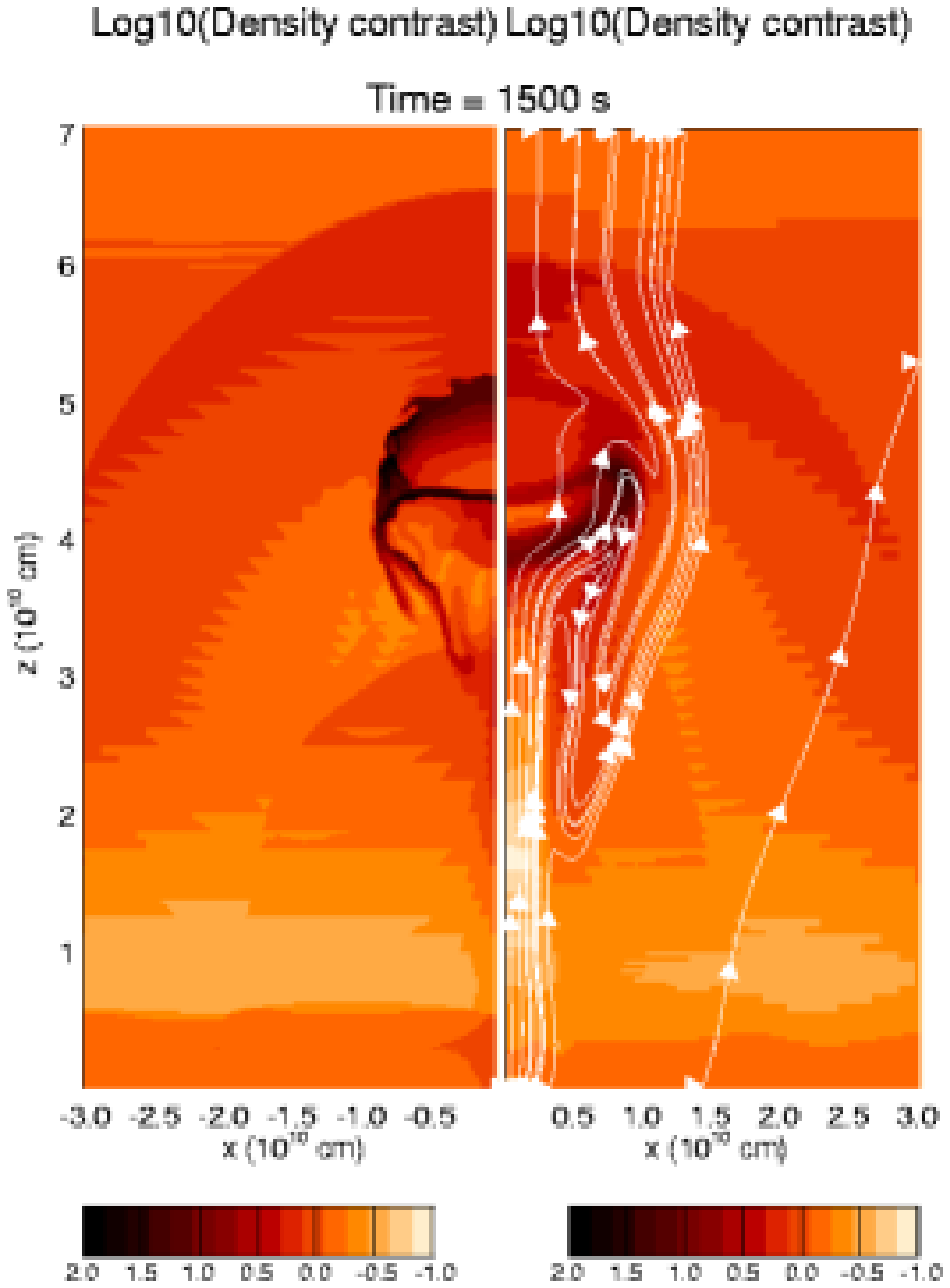}
\includegraphics[scale=0.40,clip,viewport=60 105 377 520]{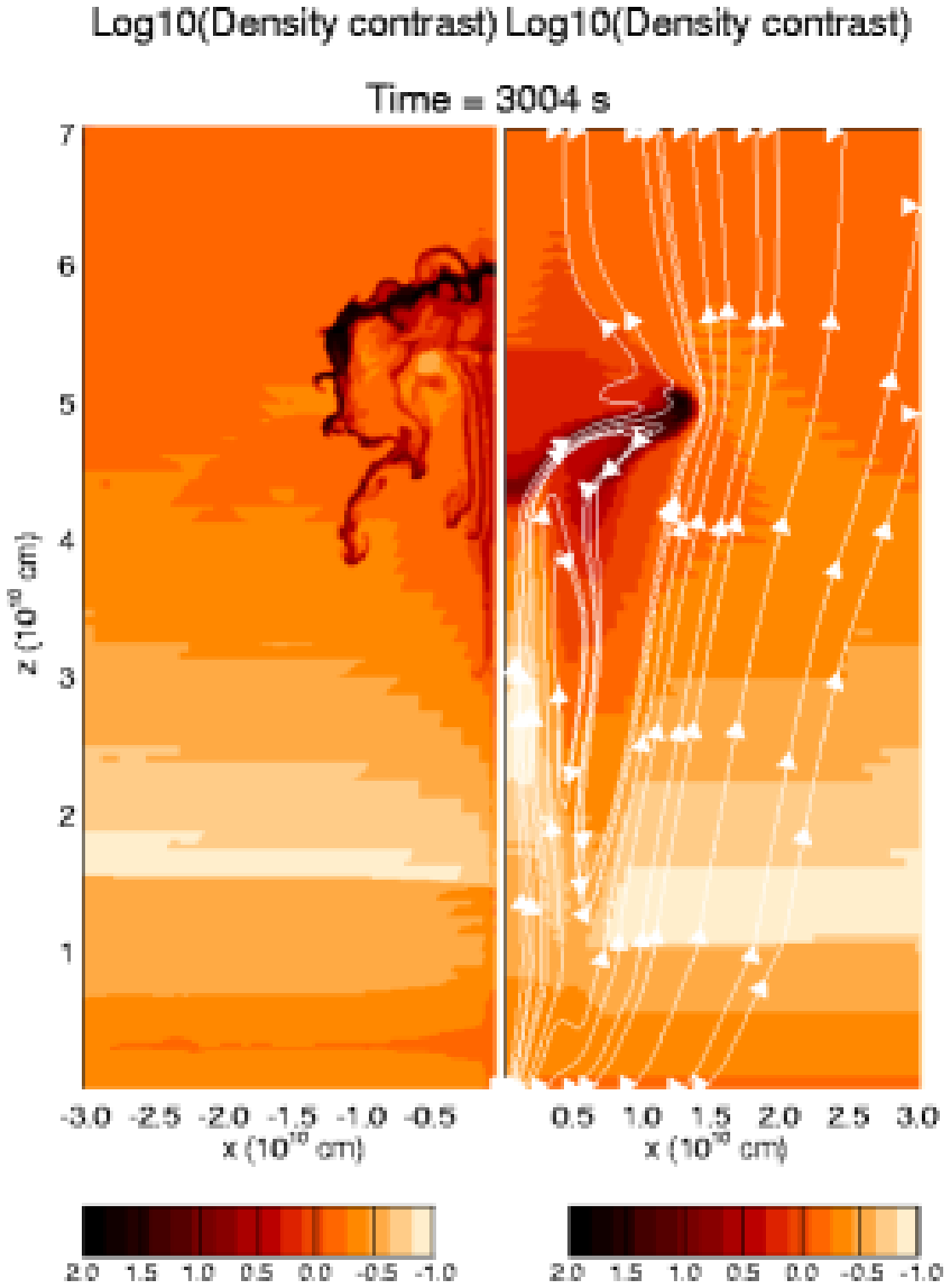}
\caption{Magnetic field lines and color maps (on the section across the cloud axis) 
of density contrast at t=500 s, 1500 s and 3000 s for the
simulations of a spherical cloud
without magnetic field ({\it left of each panel}, HS in Table~\ref{tab:sim}) and with a weak {\it open} magnetic field ({\it right of each panel}, MOHS in Table~\ref{tab:sim}).
The strength of the magnetic field ranges from $\approx0.03$ G to $\approx0.5$ G.}
\label{docyldens}
\end{figure}

We now present results for the case of a spherical cloud.
We compare the null-field evolution (the one presented with the highest detail 
in \citet{Ciaravella2001}, HS in Table~\ref{tab:sim})
with the evolution of the cloud in the weak {\it open} dipole field ($\beta\sim25$ at
the initial cloud height, MOHS in Table~\ref{tab:sim}),
performed with a fully 3-D simulation.
Figure \ref{docyldens} shows the density contrast for both the simulations.

In the HD simulation (HS), the cloud evolves into a thin dense shell-like structure, with very irregular
boundaries, due to the hydrodynamic instabilities,
that can develop in the absence of thermal conduction.
At the end of the evolution, the expansion factor of the cloud is $F_e \approx 2.7$ at $t\sim3000$ s.
Since the volume expansion of the spherical cloud scales as $r^3$, the corresponding projected expansion is
less significant than the projected expansion of the cylindrical cloud, whose volume expansion scales
as $r^2$.
In the MHD simulation (MOHS), the more effective expansion
makes the cloud apparently less conspicuous than in the case of the cylindrical cloud (Fig.\ref{do02dens}).
The expansion factor is $F_e=2.8$ at $t\sim3000$ s, $\sim4\%$ more than the HD case,
and $\sim70\%$ of the expansion of the cylindrical cloud (MOHC, see Section \ref{Magneticfield}).

\subsection{Magnetized cloud}\label{Magnetized cloud}

As mentioned in Section~\ref{model}, we consider two simulations of
magnetized clouds, one with a
weak (0.2~G) internal $B_y$-component (i.e. along the axis of the cylindrical cloud, as sketched in Fig.~\ref{fig:magn_cloud},
MOHCL (abbreviation for {\it Magnetohydrodynamic model, Open field, High $\beta$, Cylindrical cloud, Low internal magnetic field}) in Table~\ref{tab:sim},
the other with a strong (1~G) $B_y$-component (MOHCH)
in Table~\ref{tab:sim}.
We will not comment much on the latter case, because
we found that the whole cloud remains very cool all over the computed
evolution, much cooler than the temperature of formation of the relevant
UVCS lines. This case therefore seems to be unrealistic.

\begin{figure}[!htcb]
\centering
\includegraphics[scale=0.40,clip,viewport=60 125 377 540]{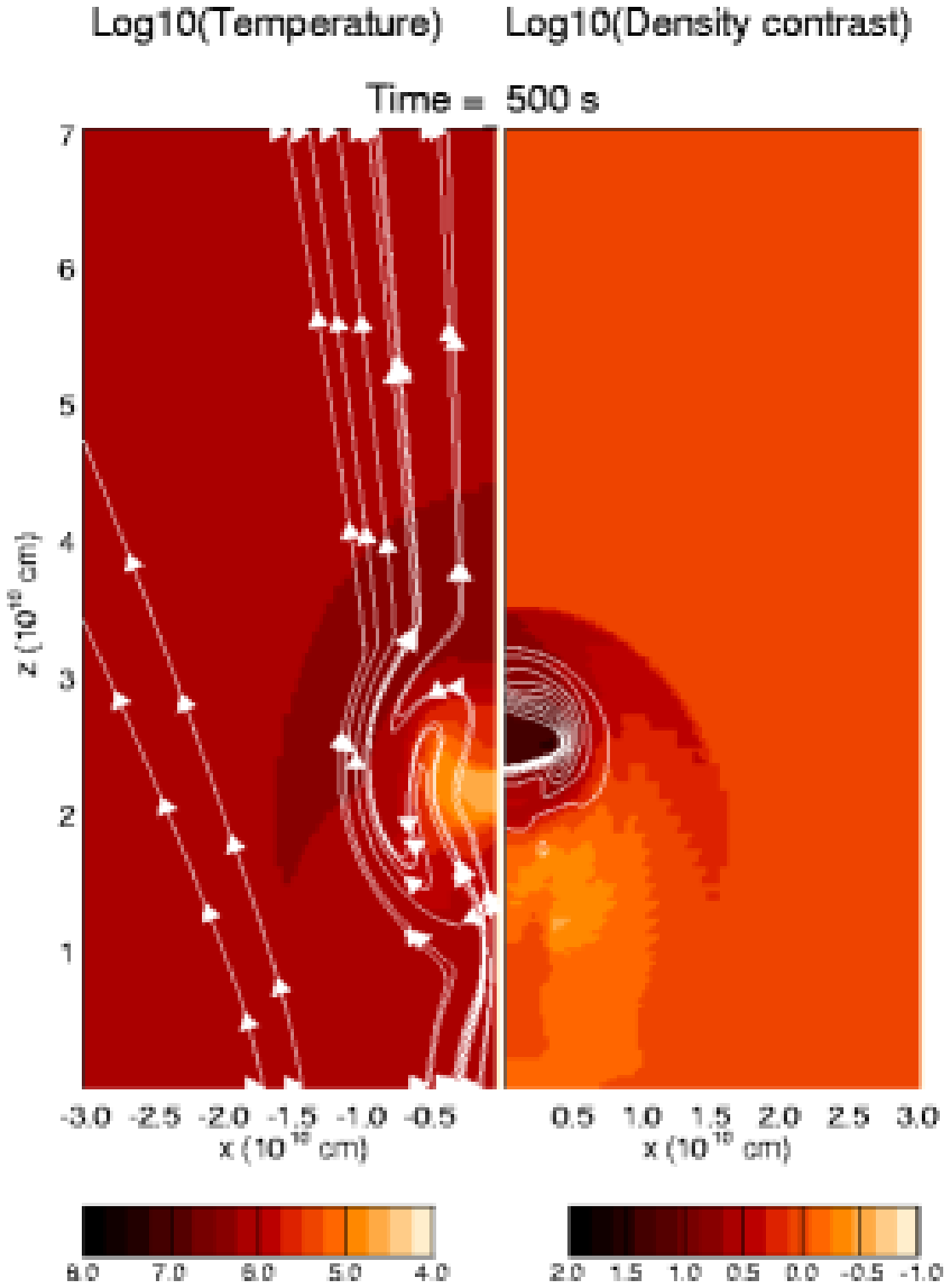}
\includegraphics[scale=0.40,clip,viewport=60 125 377 520]{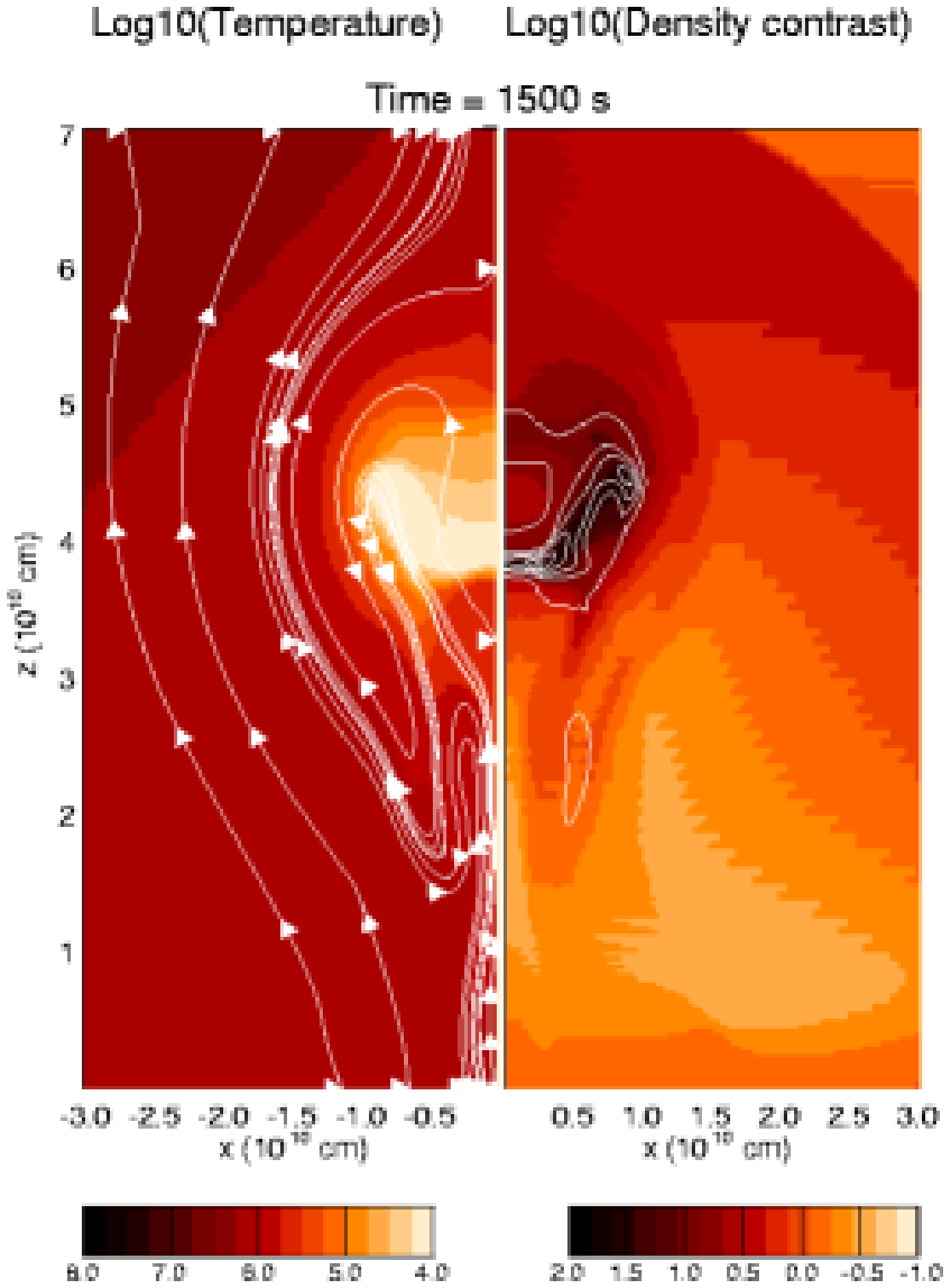}
\includegraphics[scale=0.40,clip,viewport=60 105 377 520]{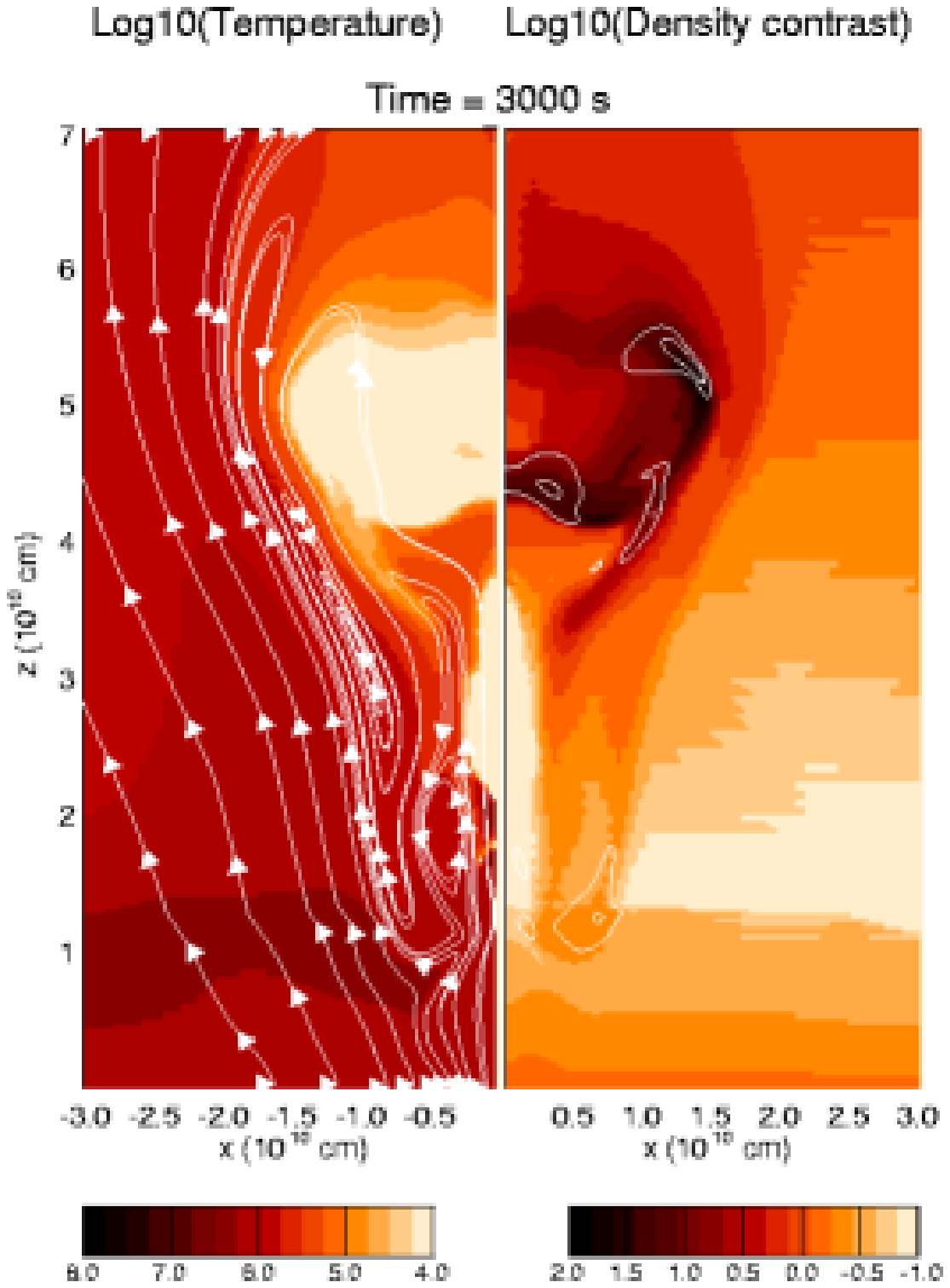}
\caption{Magnetic field lines (x-z plane, {\it left of each panel}), color maps of temperature ({\it left}) and density 
contrast ({\it right}), and contours of
the magnetic field $y$-component ({\it right}) at time t=500 s, 1500 s and 3000 s for the simulation
with a weak {\it open} ambient field and a
weak magnetic field $y$-component inside the cloud (MOHCL in Table~\ref{tab:sim}).
The contours of $B_y$ are spaced by $0.02$ G to a maximum of 0.2 G.
The strength of the magnetic field on the left of the panels ranges from $\approx0.03$ G to $\approx0.5$ G.}
\label{docmdenstemp}
\end{figure}

Figure \ref{docmdenstemp} shows magnetic field, density contrast and temperature
distributions over the cloud cross-section for the simulation with the weaker magnetic field component inside the cloud (MOHCL).
The global evolution
does not change significantly with respect to the case of no internal
magnetic field (MOHC, Section~\ref{Magneticfield}), and
the morphology of the cloud evolves similarly to
that of Figure \ref{do02dens}. Therefore, the presence of an internal
magnetic field component does not influence much the evolution of the cloud, in
spite of the initial cloud overpressure. The latter is only a small cloud
rapidly absorbed by the corona in $\tau\approx L/c_S\sim 250s$, where $L=5\times10^9$ cm
is the cloud size. The contours on the right side of the panels of Fig.~\ref{docmdenstemp} show 
that the $B_y$ component initially set up inside the cloud becomes
weaker and weaker as the cloud moves upwards and expands.

The cloud thermal insulation is very good in this simulation, as
indicated by the growing cold core, even more extended than in the other
simulations. The magnetic field component along the axis of a cylinder infinitely
extending horizontally further inhibit thermal exchanges
with the surrounding corona. In a more realistic configuration of an
elongated cloud with an internal magnetic field which bends downwards
and eventually connects to the photosphere (e.g. a flux rope),
some heat would be 
conducted to the footpoints, similar to the
1-D conduction case described in Ciaravella et al. (2001).
We estimate that the cloud
would thermalize in $\tau_c\sim1000$ s (see Eq.\ref{tcond}),
if the lenght scale of the cloud is the same as its height, say $h\sim10^{10}$ cm, above the photosphere.
The cloud expansion factor for this case of magnetized cloud is
$F_e \approx 3.6$ at $t\sim3000$~s.

\section{Discussion and conclusions}\label{discussion}

This work is devoted to studying the role of the magnetic fields in the
evolution of a fragment of a CME core traveling upwards in the high solar corona.
We address the late stage evolution of the cloud, in which it travels 
in a weakly magnetized atmosphere ($\beta \ga 1$).
Our simulations show that the evolution in the late stage could be described
better than in the early stage.
In particular, we focus on the thermal insulation of the cloud and on its
degree of expansion; both aspects were not well explained with a purely
hydrodynamic model \citep{Ciaravella2001}. In the simulations presented here,
we basically consider infinitely long cylindrical clouds, and
we first investigate which configuration (i.e. strength and topology) of the
ambient magnetic field could favor better the thermal insulation and the expansion of
the cloud simultaneously.

Before investigating the cloud enclosed by magnetic fields,
we screen out those ambient magnetic field
configurations which strongly brake the upward motion of the cloud,
i.e. provide magnetic confinement.  Since the plasma is
frozen to the field, the cloud moving upward drags the ambient magnetic
field. The strong magnetic tension then acts
against the cloud expansion and motion. Our simulations show
that the cloud is strongly braked in an atmosphere with $\beta<1$
(see Fig.\ref{dc2dens}), and is much easier to move with $\beta>1$.

In a weak ambient magnetic field ($\beta>1$), the cloud 
expands while it moves.  The evolution naturally leads to the strong thermal insulation of the
cloud because it is fast enough to drag the magnetic field so as to be
``wrapped'' by it, and to be thermally insulated from the surroundings.
For weak magnetic fields, the {\it closed} field topology yields the best thermal
insulation because the magnetic field perfectly envelops the rising cloud. 
However, for the same reason, it limits the cloud expansion and motion.

The observation supports the result of the linear expansion of a factor of 3-4 (Ciaravella et al. 2000).
In the circumstances of the cylindrical clouds, we find an expansion factor $F_e \approx 2.1$ for the
{\it closed} magnetic field simulation, $F_e=3.9$ for the {\it open} magnetic field
and an intermediate value $F_e \approx 2.9$
for the basic hydrodynamic case.
Thus, the open field favours the cloud expansion,
while the closed field does not.  
The expansion of the cloud nearly
stops at the end of the simulations because the pressure equilibrium
is reestablished between the cloud and the surroundings, and, as
\citet{Riley2004} pointed out, the magnetic tension is not important as
restoring force.

We find similar results and an expansion factor  $F_e=3.6$
for the simulation of a cylindrical cloud which carries
an initial moderate magnetic field component along its axis. A strong internal
field does not appear realistic in our conditions because it would imply an
extremely cool cloud all along its evolution, in contrast with observations. 
The effect of the internal magnetic field component is to further funnel 
heat transport and therefore reduce thermal exchanges with the surroundings.
However, we should expect thermal exchanges at the extremes of a more realistic finite cylindrical cloud.

We conclude that the {\it open} field topology is most conducive to
both the thermal insulation and a good degree of expansion, and
may therefore be the best match to observations.
This is true regardless if we consider a purely plasma cloud or
a cloud accounting also for the ejection of magnetic flux.
Several CME models already considered an {\it open} ambient magnetic
field in the outer corona in order to take into account a steady
solar wind flow  \citep{Manchester2004,Chane2005,Lugaz2005}.
Moreover, \citet{Cargill2002} considered the
model of an emerging flux rope evolving in a {\it open} magnetic field.

Our work shows that the cloud mass and shape are also important.  The
expansion factor of an initially spherical cloud increases by $\sim
4\%$ more in an ambient {\it open} field than with no field.  Instead,
for an elongated cloud the increase is of $\sim35\%$.  Thus,
although the {\it open} field induces a larger expansion regardless the
cloud geometry, this effect is more significant for an elongated
cloud than for a spherical one. This is in agreement with many
models that consider a filament eruption or a emerging flux rope as the
possible initiation or triggering of a CME \citep{Forbesisenberg1991,
Linker2003,Moore2001,Torok2005}.

We finally remark that we assume conditions of ideal MHD (except for
numerical diffusivity that gives an effective magnetic Reynolds number
$R_m \apprge 50$). It will be interesting to investigate in the future the role of the magnetic diffusion
and joule heating, which may be important locally in some areas, in which the magnetic field is found
to be enhanced during the propagation of the cloud.

\begin{acknowledgements}
The authors thank a lot Angela Ciaravella for fruitful discussion and feedback on observational aspects, and
the referee for constructive and helpful criticism. They
acknowledge support for this work from Agenzia Spaziale
Italiana, Istituto Nazionale di Astrofisica and
Ministero dell'Universit\`a e Ricerca.
The software used in this work was in part developed by the DOE-supported ASC / Alliance Center for Astrophysical Thermonuclear Flashes
at the University of Chicago,
using modules for thermal conduction and optically thin radiation built at the Osservatorio Astronomico di Palermo.
The calculation were performed on the Exadron Linux cluster at the SCAN (Sistema di Calcolo per l'Astrofisica Numerica) facility of
the Osservatorio Astronomico di Palermo and on the IBM/SP5 machine at CINECA (Bologna, Italy).
Part of the simulations were performed within a project approved in the INAF/CINECA agreement 2006-2007.
\end{acknowledgements}

\bibliographystyle{aa}
\bibliography{ref}

\end{document}